\documentclass[12pt]{article}
\usepackage{latexsym}
\usepackage{amsmath,amssymb,graphicx}
\usepackage{color,bm}
\usepackage{comment}

\def\mydate{19 Mar 2013} 
\def\ignore#1{{}}

\newcommand{\bea}{\begin{eqnarray}}
\newcommand{\eea}{\end{eqnarray}}

\makeatletter
\@addtoreset{equation}{section}
\makeatother

\newcommand{\beeq}{\begin{equation}}
\newcommand{\eneq}{\end{equation}}
\newcommand{\beqn}{\begin{eqnarray}}
\newcommand{\eeqn}{\end{eqnarray}}

\def\mybig{\displaystyle \strut }

\def\la{\raise.16ex\hbox{$\langle$}\lower.16ex\hbox{}  }
\def\ra{\raise.16ex\hbox{$\rangle$}\lower.16ex\hbox{} }
\def\go{\rightarrow}

\def\onehalf{ \hbox{$\frac{1}{2}$} }

\def\tr{{\rm tr \,}}
\def\eff{{\rm eff}}

\def\cF{{\cal F}}

\def\SM{{\rm SM}}

\def\diag{{\rm diag ~}}

\def\KK{{\rm KK}}

\def\psibar{ \psi \kern-.65em\raise.6em\hbox{$-$} }
\def\psibarl{ \psi \kern-.65em\raise.6em\hbox{$-$} \lower.6em\hbox{} }

\def\Psibar{ {\overline{\Psi}} }

\def\myfrac#1#2{{\mybig #1\over \mybig #2}}

\begin{document}

\thispagestyle{empty}

{\small \noindent \mydate    \hfill OU-HET 775/2013}


\vspace{4.0cm}

\baselineskip=35pt plus 1pt minus 1pt

\begin{center}
{\Large \bf 
Novel universality and  Higgs decay $H \go \gamma \gamma, gg$\\
in the $SO(5) \times  U(1)$ gauge-Higgs unification}
\end{center}

\vspace{2.0cm}
\baselineskip=20pt plus 1pt minus 1pt

\begin{center}
{\bf
Shuichiro Funatsu, Hisaki Hatanaka, Yutaka Hosotani,\\
Yuta Orikasa  and Takuya Shimotani
}


{\small \it Department of Physics, 
Osaka University, 
Toyonaka, Osaka 560-0043, 
Japan} \\
\end{center}


\vskip 2.5cm
\baselineskip=20pt plus 1pt minus 1pt

\begin{abstract}
The $SO(5) \times U(1)$  gauge-Higgs unification in the Randall-Sundrum warped space
with the  Higgs boson mass $m_H=126\,$GeV is constructed.
An  universal relation  is found between the Kaluza-Klein (KK) mass scale 
$m_\KK$ and the Aharonov-Bohm (AB) phase $\theta_H$ in the fifth dimension; 
$m_\KK \sim 1350\,{\rm GeV}/(\sin \theta_H)^{0.787}$.
The cubic and quartic self-couplings of the Higgs boson become smaller than those
in the standard model (SM), having universal dependence on $\theta_H$.
The decay rates $H \go \gamma \gamma, gg$ are evaluated by summing  contributions
from KK towers.  
Corrections coming from KK excited states are finite and 
about 0.2\% (2\%) for $\theta_H= 0.12 \, (0.36)$, branching fractions of 
various decay modes of the Higgs boson remaining nearly the same as in the SM.
The signal strengths of the Higgs decay modes relative to the SM are 
$\sim \cos^2\theta_H$.
The mass of the first KK $Z$ is predicted  to be  $5.9 \, (2.4)\, $TeV
for $\theta_H= 0.12 \, (0.36)$.
We also point out the possible enhancement of $\Gamma(H\to\gamma\gamma)$ due to the large $U(1)_X$ charge of new fermion multiplets.
\end{abstract}



\newpage

\baselineskip=20pt plus 1pt minus 1pt

With the discovery of a Higgs-like boson at LHC  \cite{atlas1, cms1} it is emergent to pin down 
its properties to see if it is the Higgs boson in the standard model (SM).
The mechanism of electroweak (EW) symmetry breaking is at issue.
It is not clear if the EW symmetry is spontaneously
broken in a way described in the SM. 
The mass of the discovered boson is about 126$\,$GeV.  Its couplings
to other fields, however,  may or may not be the same as in the SM. The excess in the
decay mode $H \go \gamma \gamma$ has been reported, 
though more data are necessary for the issue to be settled. \cite{atlas2, cms2}

Many alternative mechanisms for the EW symmetry breaking  have been proposed
with new physics beyond the SM.  
Supersymmetry with a light Higgs boson has been a popular scenario in the past, 
though no evidence has been found so far.   
It has been  discussed that the value $m_H = 126\,$GeV can lead to the  direct 
connection to physics at the Planck scale through the vacuum stability of the 
SM or conformality.\cite{meta-stable}-\cite{Orikasa}
Many scenarios have been proposed to account for the apparent excess rate
for the Higgs decay to two photons at LHC.\cite{ued}-\cite{ext-higgs}

The gauge-Higgs unification scenario is one of the models with new physics at the 
TeV scale, in which the 4D Higgs boson is identified with the zero mode of 
the extra-dimensional  component of the gauge fields.\cite{YH1}-\cite{Hatanaka1998}
In this paper we show that the value  of the Higgs boson mass $m_H = 126\,$GeV 
has profound implications in the gauge-Higgs unification.  
We evaluate the decay rates $H \go \gamma \gamma, gg$ 
by summing contributions from all Kaluza-Klein (KK) excited states of
the $W$ boson and fermions in the internal loops. Surprisingly there arises no
divergence associated with the infinite sum, thanks to destructive interference in the 
amplitude.  The corrections to the decay rates
$H \go \gamma \gamma, gg$ are finite and small, being independent of a cutoff scale.
With $m_H = 126\,$GeV as an input, the deviation of the branching fractions of the $H$ decay
from the values in the SM is  found  to be 2\% or less.  

In the  $SO(5) \times U(1)$ gauge-Higgs unification model \cite{ACP}
the 4D neutral Higgs field appears as  4D fluctuations of the AB, or Wilson line,  
phase $\theta_H$  along the fifth dimension in the Randall-Sundrum (RS) warped space-time. 
In the minimal   model with quark and lepton multiplets in the vector representation of $SO(5)$ 
\cite{HOOS, HNU}
the effective potential $V_\eff (\theta_H)$ is minimized at $\theta_H = \pm \onehalf \pi$,
where the Higgs boson becomes absolutely stable.\cite{HTU1}
This is due to the emergence of the $H$ parity invariance.\cite{HKT}
To have an unstable Higgs boson with a mass $m_H = 126\,$GeV
the model need to be modified by breaking the $H$ parity.  
Further in the minimal model the consistency with the electroweak precision 
measurements requires a large warp factor $z_L$, which typically leads to 
a larger value $m_H \sim 135\,$GeV.\cite{HTU2}
The Higgs mass can be made smaller in the supersymmetric version of 
the model.\cite{Hatanaka2012}

To solve these problems we introduce $n_F$ fermion multiplets,  $\Psi_F$, 
in the spinor representation of $SO(5)$
in the model specified in Ref.~\cite{HNU}.    
The metric of  the RS  is given by
$ds^2 = e^{-2 \sigma(y)} \eta_{\mu\nu} dx^\mu dx^\nu + dy^2$
where $\sigma(y) = k |y|$ for $|y| \le L$ and $\sigma(y+2L) = \sigma(y)$.  
The warp factor is $z_L = e^{kL} \gg 1$.
$\Psi_F$ satisfies the boundary conditions $\Psi_F(x,-y) = \gamma_5 P \Psi_F(x, y)$ and
$\Psi_F(x, L-y) = - \gamma_5 P \Psi_F(x, L+y)$ where $P = \diag (1,1,-1,-1)$ acts on
$SO(5)$ spinor indices.  Then the mass spectrum $m_n = k \lambda_n$ of the KK tower 
of $\Psi_F$ is determined by 
$S_L (1; \lambda_n, c_F) S_R (1; \lambda_n, c_F) + \cos^2 \onehalf \theta_H = 0$.
Here $S_{L,R} (z; \lambda, c) = \mp \onehalf \pi \lambda \sqrt{z z_L} 
F_{c\pm (1/2), c\pm (1/2)} (\lambda z, \lambda z_L)$,  where the upper (bottom)
sign refers to $L$ ($R$). 
$F_{\alpha, \beta} (u,v) = J_\alpha (u) Y_\beta (v) - Y_\alpha (u) J_\beta (v)$ 
where $J_\alpha$, $Y_\alpha$ are Bessel functions.
It has been shown that all 4D anomalies in the model of Ref.~\cite{HNU} cancel.
This property is not spoiled by the addition of $\Psi_F$ multiplets, as 4D
fermions of $\Psi_F$ are vectorlike.

The  effective potential $V_\eff (\theta_H)$ is cast in a simple form of an integral.
The relevant part of $V_\eff (\theta_H)$ is given by 
\beqn
&&\hskip -1.cm
V_\eff (\theta_H; \xi, c_t, c_F, n_F, k, z_L) =
 2 (3-\xi^2) I[Q_W] + (3-\xi^2) I[Q_Z] + 3 \xi^2 I[Q_S]  \cr
\noalign{\kern 10pt}
&&\hskip 3.6cm
-12  \big\{ I[Q_{\rm top}] + I[ Q_{\rm bottom}] \big\} - 8 n_F  I[Q_F] ~, \cr
\noalign{\kern 10pt}
&&\hskip -1.cm
I[Q(q; \theta_H)] = \frac{(k z_L^{-1})^4}{(4\pi)^2}
\int_0^\infty dq \, q^3 \ln \{ 1 + Q(q; \theta_H) \} ~, \cr
\noalign{\kern 10pt}
&&\hskip -1.cm
Q_W =  \cos^2 \theta_W  Q_Z  = \onehalf Q_S = \onehalf Q_0[q; \onehalf] \sin^2 \theta_H ~, \cr
\noalign{\kern 10pt}
&&\hskip -1.cm
Q_{\rm top} = \frac{Q_{\rm bottom} }{r_t} 
= \frac{ Q_0[q; c_t]}{2(1+ r_t)} \, \sin^2 \theta_H ~,~~ 
Q_F =  Q_0[q; c_F] \cos^2 \onehalf \theta_H ~, \cr
\noalign{\kern 10pt}
&&\hskip -1.cm
Q_0[q; c] =  \frac{z_L}{q^2 \hat F_{c - \onehalf,c - \onehalf} (qz_L^{-1}, q) 
              \hat F_{c + \onehalf, c + \onehalf} (qz_L^{-1}, q) }  ~.
\label{effV1}
\eeqn
Here $\hat F_{\alpha ,\beta} (u, v) = I_\alpha (u) K_\beta(v) 
- e^{- i (\alpha - \beta) \pi}  K_\alpha (u) I_\beta(v)$, where $I_\alpha$, $K_\alpha$ are 
modified Bessel functions.
$\xi$ is a gauge parameter in the generalized R$_\xi$ gauge introduced in Ref.~\cite{HNU}.
The formula for $V_\eff $ in the $\xi=1$ gauge without the  $I[Q_F]$ term has been given 
in Refs.\ \cite{HOOS} and \cite{HTU1}.  $c_t$ and $c_F$ are the bulk mass parameters
for the top-bottom multiplets and $\Psi_F$, respectively. $r_t \sim (m_b/m_t)^2$ 
where $m_b$ and $m_t$ are the masses of the bottom and top quark.
$V_\eff (-\theta_H) = V_\eff (\theta_H)$.  Further
in the absence of $I[Q_F]$, $V_\eff$ 
has symmetry $V_\eff (\onehalf \pi + \theta_H ) = V_\eff (\onehalf \pi - \theta_H ) $,
representing the $H$ parity invariance.  The $I[Q_F]$ term breaks this symmetry.
The contributions from light quarks and leptons are negligible.

In the pure gauge theory without fermions $V_\eff $ is minimized at $\theta_H=0, \pi$ 
where the EW symmetry remains unbroken.  The top quark contribution has  minima
at $\theta_H = \pm \onehalf \pi$, dominating over the gauge field contribution.
The fermion $\Psi_F$ shifts the minima toward $\theta_H =0$.  The minimum at
$0 <  | \theta_H  | < \onehalf \pi$ gives desired phenomenology.  
The number of the fermion multiplets $\Psi_F$, $n_F$, affects the shape of $V_\eff $
significantly.  It will be shown below that the resulting physics, however,  is almost independent
of $n_F$.
The mass of the Higgs boson, $m_H$, is given by
\beeq
m_H^2 = \frac{1}{f_H^2} \frac{d^2 V_\eff}{d \theta_H^2} \bigg|_{\rm min} , ~~
f_H = \frac{2}{g_w} \sqrt{ \frac{k}{L(z_L^2 -1)}}
\eneq
where the second derivative of $V_\eff$ is evaluated at the minimum of $V_\eff$, and 
$g_w$ is the 4D weak $SU(2)_L$ coupling.  The experimental data dictate
$m_H \sim 126\,$GeV.  

\begin{figure}[htb]
\begin{center}
\includegraphics[height=4.5cm]{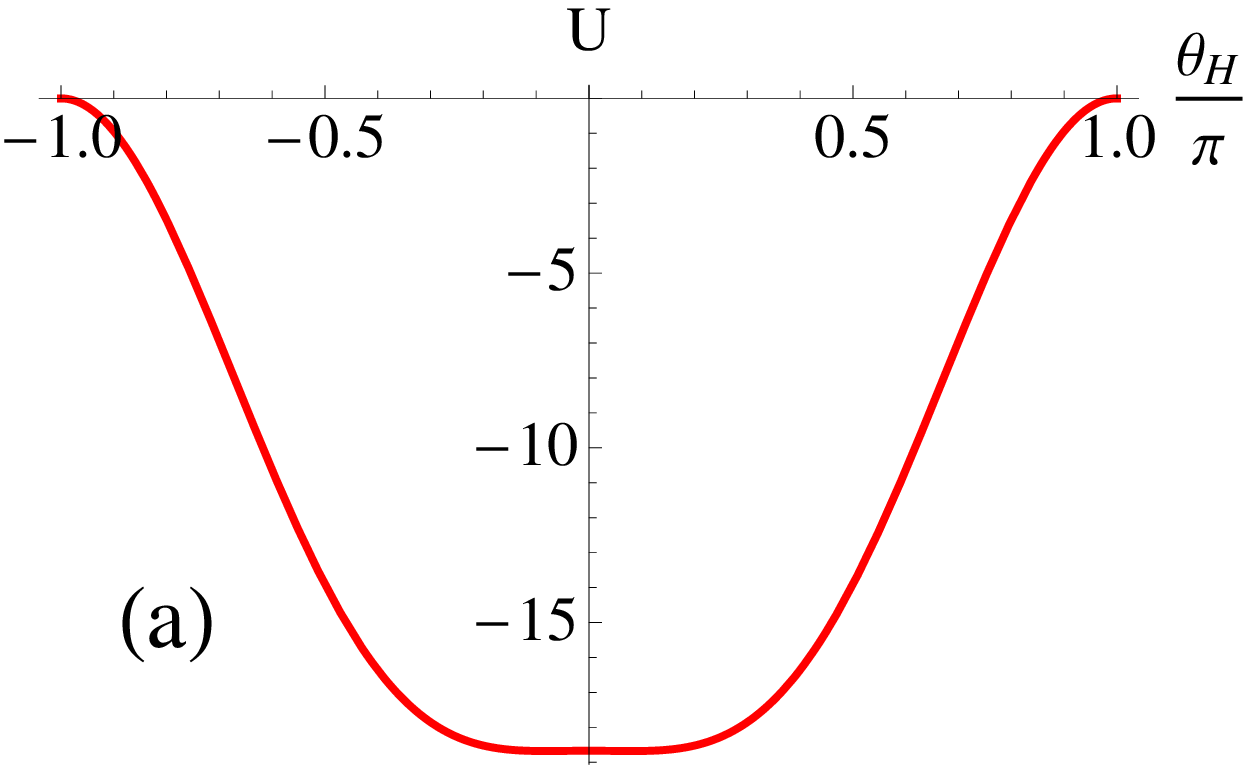}
\includegraphics[height=4.5cm]{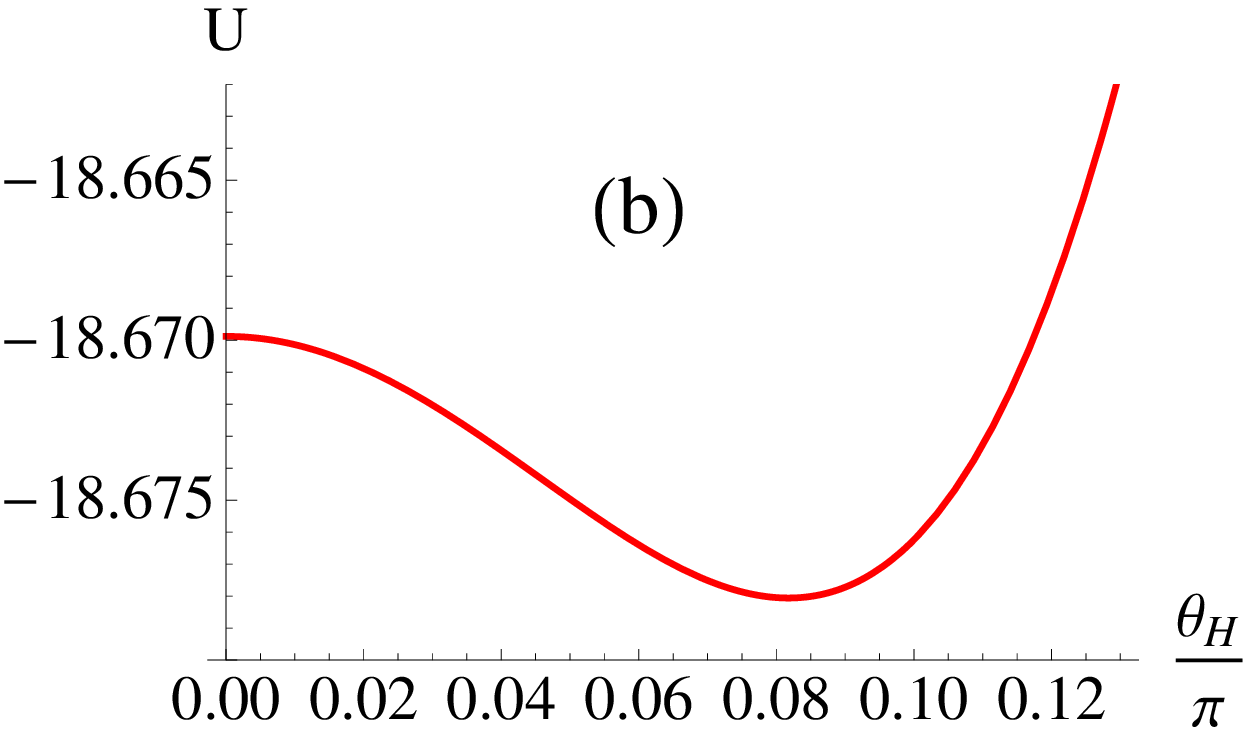}\\
\caption{$V_\eff(\theta_H)$ for $n_F=3$, $c_F=0.353$, $z_L= 10^7$ and $\xi=1$.
$U = (4\pi)^2 (kz_L^{-1})^{-4} V_\eff$ is plotted.
The minimum is located at $\theta_H   = \pm 0.082 \, \pi =  \pm 0.258$.
(a): $-\pi \le \theta_H \le \pi$, (b): $0 \le \theta_H \le 0.13\, \pi$.
}
\label{figVeff}
\end{center}
\end{figure}

The parameters of the model are specified in the following manner.
Pick values for $n_F$ and $z_L$.  
The parameters, $k$, two gauge coupling constants associated with $SO(5) \times U(1)$, 
$c_t$, $r_t$, and $c_F$ are
self-consistently determined such that at the minimum $\theta_H$ of $V_\eff$, 
$m_Z$, $\sin^2 \theta_W$, $\alpha (m_Z)$,  $m_t$,   $m_b$ and $m_H=126\,$GeV
are reproduced.  
\ignore{
Suppose that $V_\eff$ is minimized at $\theta_H = \theta_H^{(1)}$.
Then from the values of $m_Z$, $\sin^2 \theta_W$, $\alpha (m_Z)$, $m_t$, and $m_b$
the parameters of the model $k$, two gauge coupling constants, $c_t$, and $r_t$
are determined. With an additional input $c_F$, $V_\eff (\theta_H)$ in (\ref{effV1}) is fixed,
which in general has a minimum at $\theta_H^{(2)} (c_F)$ and yields $m_H (c_F, \theta_H^{(1)})$.
We determine $\theta_H^{(1)}$ and $c_F$  such that $\theta_H^{(2)} (c_F) = \theta_H^{(1)}$
and $m_H (c_F, \theta_H^{(1)}) = 126\, $GeV.  
}
We note that all of $k$,  $c_t$,   $r_t$ and $c_F$ implicitly depend on $\theta_H$ as well.
The KK mass scale is given by $m_\KK = \pi k z_L^{-1}$.  
Hence $m_\KK$ and $\theta_H$, for instance, are determined as functions of $n_F$ and $z_L$.
$V_\eff (\theta_H)$ for $n_F=3$ and $z_L= 10^7$ is displayed in Fig.~\ref{figVeff}.

The AB phase $\theta_H$ is the key parameter in gauge-Higgs unification, which controls
the couplings of the Higgs boson to other fields.  
In Table \ref{table1} the values for $z_L$, $\theta_H$, $m_\KK$, $k$, $c_t$, $c_F$,
$m_{F^{(1)}}$ and $m_{Z^{(1)}}$ are summarized  for $n_F=3$,
where $m_{F^{(1)}}$ is the mass of the lowest mode in the KK tower of $\Psi_F$
and $m_{Z^{(1)}}$ is the mass of the first KK $Z$ boson.  
As $z_L$ is decreased, $\theta_H$ becomes smaller whereas $m_\KK$ becomes larger.
There appears a critical value for $z_L$ below which $m_H=126\,$GeV cannot be
realized.

\begin{table}[htdp]
\caption{Values of the various quantities determined from $m_H=126\,$GeV 
with given $z_L$ for $n_F=3$.
Universal relations among $\theta_H$, $m_\KK$ and $m_{Z^{(1)}}$, 
independent of $n_F$,  are observed.  See the text.}
\begin{center}
\begin{tabular}{|c|ccccccc|}
\hline
$z_L$  & $\theta_H$ & $m_{\mbox {{\tiny KK}}}$ & $k$ 
& $c_t$ & $c_F$ &$m_{F^{(1)}}$  &$ m_{Z^{(1)}} $ \\
& & {\small (TeV)} &{\small (GeV)}
&& &{\small (TeV)} &{\small (TeV)}\\
\hline
$10^{12}$  & $1.02~$ & $1.54$ & $4.90\times 10^{14}$& $0.413$& $0.476$ &$0.155$&$ 1.19 $\\
$10^{11}$  & $ 0.805$ & $ 1.75$ & $5.56\times 10^{13}$ & $0.403$& $ 0.454$&$0.232$&$ 1.36 $\\
$10^{10}$ & $ 0.632$ & $ 2.03$ & $ 6.47\times 10^{12}$ & $ 0.391$ & $0.433$&$0.329$&$ 1.59 $\\
$10^9$  & $0.485$ & $ 2.45$ & $7.79\times 10^{11}$& $ 0.376$& $0.411$ &$0.465$&$ 1.93 $\\
$10^8$   & $0.360$ & $ 3.05$ & $ 9.72\times 10^{10}$ & $ 0.357$&$ 0.385$&$0.668$&$ 2.41 $\\
$10^7$  & $0.258$ & $ 3.95$ & $ 1.26\times 10^{10}$& $0.330$& $ 0.353$ &$0.993$&$3.15  $\\
$10^6$  & $ 0.177$ & $5.30$ & $ 1.69\times 10^{9}$ & $ 0.296$& $ 0.309$&$1.54~$&$ 4.25 $\\
$10^5$ & $0.117$ & $ 7.29$ & $ 2.32\times 10^{8}$& $0.227$& $ 0.235$  &$2.53~$&$ 5.91 $\\
$2\times10^4$ & $0.086$ & $9.21$ & $5.87\times 10^{7}$ & $ 0.137$& $ 0.127$& $3.88~$&$ 7.54$\\
\hline
\end{tabular}
\end{center}
\label{table1}
\end{table}

\begin{figure}[htb]
\begin{center}
\includegraphics[height=5.5cm]{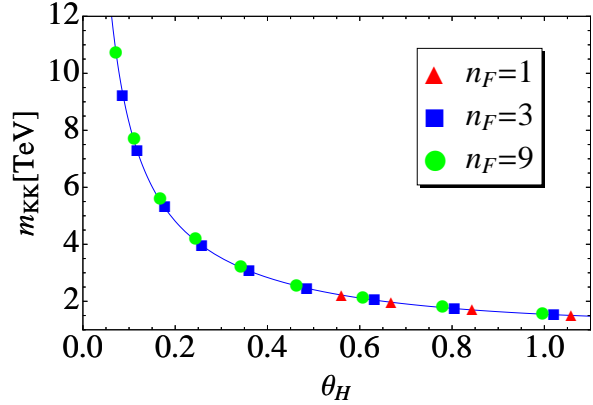}\\
\caption{The relation between $\theta_H$ and $m_\KK$ for $\xi = 1$. 
Triangles, squares, and circles are for $n_F=1, 3$ and 9, respectively.
The solid curve represents the universal relation (\ref{mKKtheta1}).
}
\label{mKK1}
\end{center}
\end{figure}

Both $\theta_H$ and $m_\KK$ are  physical quantities.  They are functions of $n_F$ 
and $z_L$.   The relation between them are plotted in Fig.~\ref{mKK1} for various 
values of  $n_F$ and $z_L$.   
As the number of the extra fermions $n_F$ is increased, the location of the minimum 
of $V_\eff$ is shifted toward the origin.
Nevertheless the relation between $\theta_H$ and $m_\KK$ remains universal.  
It is approximately given by
\beeq
m_\KK \sim \myfrac{1350 \, {\rm GeV}}{( \sin \theta_H )^{0.787}} ~,
\label{mKKtheta1}
\eneq
irrespective of $n_F$ and $z_L$.  We note that 
$m_Z \sim m_\KK |\sin \theta_H|/( \pi \cos \theta_W  \sqrt{kL} )$, in which 
$\theta_H$ and $kL = \ln z_L$ are not independent, once $m_H$ is fixed.
There must be an underlying reason for the universality relation (\ref{mKKtheta1}), 
which remains as a mystery and is left for future investigation.  
The relation between $\theta_H$ and $m_\KK$,  
with $m_\KK > 3\,$TeV for the  consistency with low energy data, 
implies that $\theta_H < 0.3$, which also satisfies 
the $S$ parameter constraint \cite{ACP} and the tree-level unitarity 
constraint \cite{unitarity}.
For $\theta_H = 0.1 \sim 0.3$, $m_\KK$ is  predicted to be around $3 \sim 7\,$TeV,   
in a region which can be explored at LHC in the coming years.
We have also checked that the $m_\KK$-$\theta_H$ relation in the  $\xi=0$ gauge
is almost the same as in the $\xi=1$ gauge.

\begin{figure}[htb]
\begin{center}
\includegraphics[height=4.8cm]{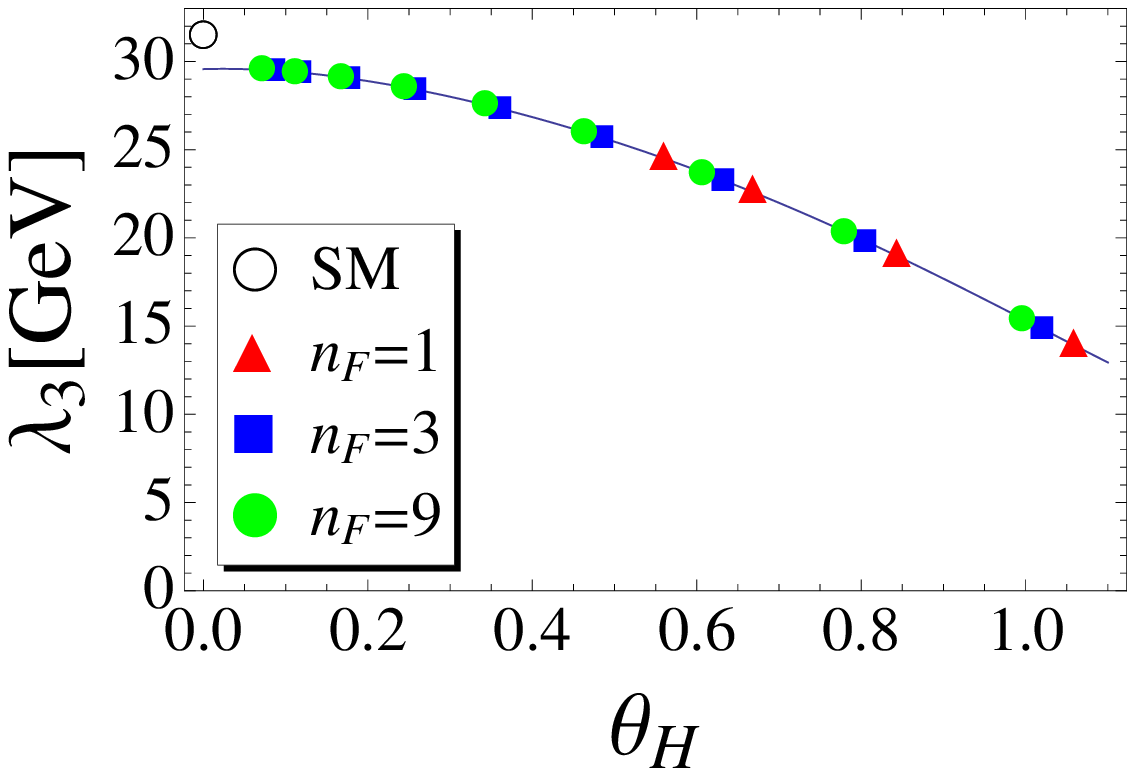}
\includegraphics[height=4.8cm]{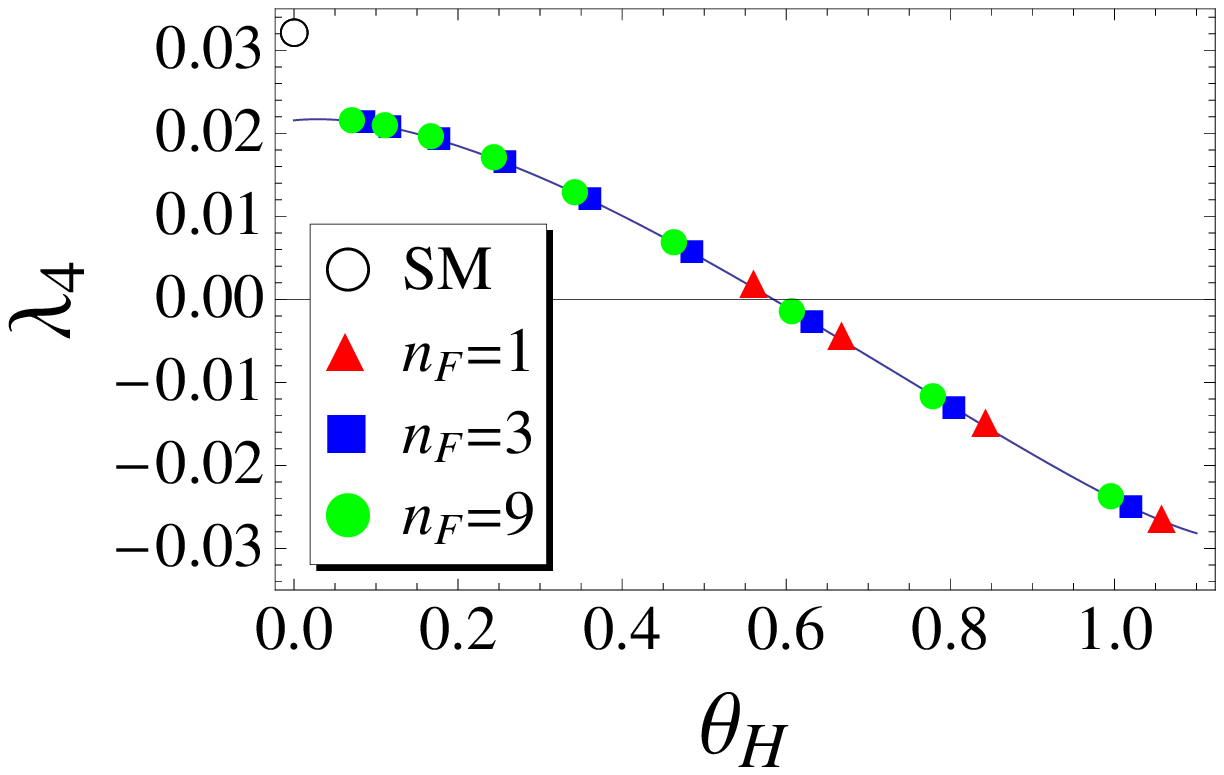}\\
\caption{The  cubic coupling $\lambda_3$ and  quartic coupling $\lambda_4$
of the Higgs boson
are plotted against $\theta_H$ for $n_F=1, 3$ and 9. 
The universal relations between $(\lambda_3, \lambda_4)$ and $\theta_H$, 
independent of  $n_F$,  are seen.  
The fitting curves for $\lambda_3$ and $\lambda_4$ are given by 
$\lambda_3 = 5.28 + 21.1 \cos \theta_H + 3.20 \cos 2 \theta_H$ and
$\lambda_4 = 0.0256 - 0.0537 \cos\theta_H + 0.0495 \cos 2\theta_H$, respectively.
The SM values are $\lambda_3   = 31.5\,$GeV  and $\lambda_4 = 0.0320$.\cite{Degrassi2012}
}
\label{Hselfcoupling}
\end{center}
\end{figure}

The gauge-Higgs unification model has one parameter, $\theta_H$, to be determined
from experiments.  With $\theta_H$ fixed, all physical quantities are evaluated.
By expanding $V_\eff (\theta_H + (H/f_H))$ in a power series in $H$ around
the minimum, one finds $\lambda_n H^n$ couplings.  These couplings $\lambda_3$ 
and $\lambda_4$
\ignore{relative to those in the SM} 
are plotted in Fig.~\ref{Hselfcoupling} for $n_F = 1, 3$ and 9.
The couplings are smaller than those in the SM.  
For large $\theta_H > 0.55$, $\lambda_4$ becomes negative though $V_\eff$ is 
bounded from below.
It is seen that the relations $\lambda_3 (\theta_H)$ and $\lambda_4 (\theta_H)$ 
are also universal and independent of $n_F$, once $m_H=126\,$GeV is fixed.

The Higgs couplings to $W$, $Z$, quarks/leptons and their KK excited states are
determined.  All of the 3 point couplings of SM particles to $H$ at the tree level are  
suppressed by a common factor $\cos \theta_H$.\cite{HS2}-\cite{Hasegawa}  
It is necessary to find these 3 point couplings of KK states for evaluating 
the 1-loop processes such as $gg \go H$ and $H \go \gamma \gamma, \,  gg$.

Let us first consider the process $H \go \gamma \gamma$.  It proceeds through one-loop
diagrams.  All charged particles with non-vanishing Higgs couplings contribute.
The dominant contributions come from the $W$ boson, top quark, and their KK towers.
The extra fermion $\Psi_F$ also contributes.  The decay rate is given by
 \cite{Ellis1975, Shifman1979, HiggsHunter}
\beqn
&&\hskip -1.cm
\Gamma ( H \go \gamma \gamma ) = \frac{\alpha^2 g_w^2}{1024 \pi^3}
\frac{m_H^3}{m_W^2} \, \Big| \cF_W + \frac{4}{3} \cF_{\rm top}
+ \left(2 (Q_X^{(F)})^2 + \onehalf\right)  n_F \cF_{F} \Big|^2 ~,  \cr
\noalign{\kern 10pt}
&&\hskip -1.cm
\cF_W = \sum_{n=0}^\infty \frac{g_{HW^{(n)} W^{(n)} }}{g_w \, m_W} 
\frac{m_W^2}{m_{W^{(n)}}^2} \,  F_1 ( \tau_{W^{(n)}} ) ~, \cr
\noalign{\kern 10pt}
&&\hskip -1.cm
\cF_{\rm top} = \sum_{n=0}^\infty \frac{y_{t^{(n)}}}{y_t^\SM} \frac{m_t}{m_{t^{(n)}}} \, 
F_{1/2} ( \tau_{t^{(n)}} ) ~, \cr
\noalign{\kern 10pt}
&&\hskip -1.cm
\cF_F = \sum_{n=1}^\infty \frac{y_{F^{(n)}}}{y_t^\SM} \frac{m_t}{m_{F^{(n)}}} \, 
F_{1/2} ( \tau_{F^{(n)}} ) ~, 
\label{Hdecay1}
\eeqn
where $W^{(0)} = W$, $t^{(0)} =t$, $\tau_{a} = 4 m_{a}^2/m_H^2$ and 
the functions $F_1(\tau)$ and $F_{1/2}(\tau)$ are defined in Ref.~\cite{HiggsHunter}.
$Q_X^{(F)}$ is the $U(1)_X$ charge of $\Psi_F$.
$y_t^\SM$ denotes the top Yukawa coupling in the SM.
Note that $F_1(\tau) \go 7$ and $F_{1/2} (\tau) \go - \frac{4}{3}$ for $\tau \go \infty$. 
The extra fermion multiplet $\Psi_F$ contains particles with electric charges $(Q_X^{(F)}\pm \onehalf) e$.
It will be seen below that the contribution $\cF_F$
is small for  $\theta_H < 0.5$.
The $H W^{(n)} W^{(n) \dagger}$ coupling $g_{HW^{(n)} W^{(n)} }$ and 
the Yukawa couplings $y_{t^{(n)}}$ and $y_{F^{(n)}}$ are unambiguously determined 
in the gauge-Higgs unification.
The infinite sums in (\ref{Hdecay1}) turn out   finite.
The expression for $\cF_W$ corresponds to the amplitude in the unitary gauge.
It has been shown in Ref.~\cite{Marciano} that the correct amplitude is reproduced
in the unitary gauge  in the SM.  

In the gauge-Higgs unification the $H W^{(n)} W^{(n) \dagger}$ and Yukawa couplings result from
the $\tr F_{\mu 5} F^{\mu 5}$ term and $\Psibar \Gamma^5 A_5 \Psi$ terms  in the action,
where the vector potential $A_5$ contains the 4D Higgs  field.  To good approximation
$g_{HW W} \sim g_{HW W}^{\SM} \cos \theta_H = g_w m_W \cos \theta_H $  and
$y_t \sim y_t^{\SM} \cos \theta_H$.  

\begin{figure}[bht]
\begin{center}
\includegraphics[height=5.5cm]{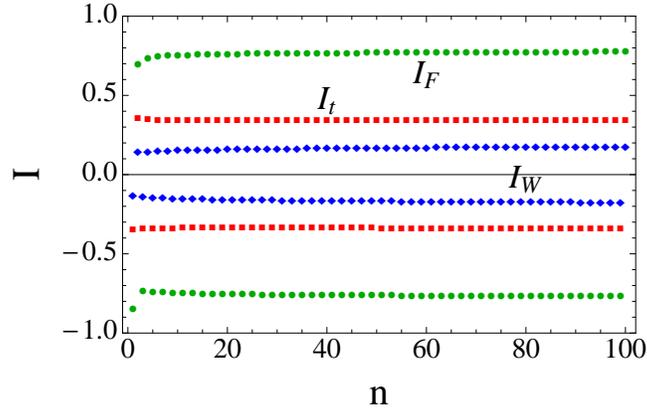}\\
\caption{The ratios 
$I_{W^{(n)}} = g_{HW^{(n)} W^{(n)}} / g_w m_{W^{(n)}} \cos\theta_H$ in (\ref{HWW1}), 
$I_{t^{(n)}} = y_{t^{(n)}} / y_t^\SM \cos \theta_H$   in (\ref{Htt1})  and
$I_{F^{(n)}} = y_{F^{(n)}} / y_t^\SM \sin \onehalf \theta_H$  in (\ref{HFF1}) are plotted
for $n_F=3$ and   $\theta_H= 0.360$ ($z_L=10^8$) in the range $1 \le n \le 100$.
($\Box$: the top quark tower, $\Diamond$: the $W$ tower, {\small $\bigcirc$}: the $\Psi_F$ tower)
$I_{W^{(0)}} = 1.004$ and $I_{t^{(0)}} = 1.012$.
The sign of $g_{HW^{(n)} W^{(n)}}$, $y_{t^{(n)}}$ and $y_{F^{(n)}}$ alternates
as $n$ increases. $I_{W^{(1)}}, I_{t^{(1)}}, I_{F^{(1)}} < 0$.
}
\label{HWWyukawa}
\end{center}
\end{figure}

One finds that
\beqn
&&\hskip -1.cm
I_{W^{(n)}} = 
\frac{g_{HW^{(n)} W^{(n)}} }{ g_w m_{W^{(n)}} \cos \theta_H }
= - \sqrt{ kL (z_L^2-1) } ~ \frac{\sin\theta_H}{N_{W^{(n)}}} \, 
\frac{C(1; \lambda_{W^{(n)}})}{S(1; \lambda_{W^{(n)}})} ~, \cr
\noalign{\kern 10pt}
&&\hskip -1.cm
N_{W^{(n)}} = \int_1^{z_L} \frac{dz}{z} 
\Big\{ (1+ \cos^2 \theta_H) C(z; \lambda_{W^{(n)}})^2
+ \sin^2 \theta_H \hat S(z; \lambda_{W^{(n)}})^2 \Big\} ~, \cr
\noalign{\kern 10pt}
&&\hskip -.5cm
C(z; \lambda) = \frac{\pi}{2} \lambda z z_L F_{1,0}(\lambda z, \lambda z_L) ~, \cr
\noalign{\kern 10pt}
&&\hskip -.5cm
S(z; \lambda) = - \frac{\pi}{2} \lambda z  F_{1,1}(\lambda z, \lambda z_L) ~, ~~
\hat S(z; \lambda) = \frac{C(1; \lambda) }{S(1; \lambda) } S(z; \lambda) ~.
\label{HWW1}
\eeqn
We note that $S(1; \lambda_{W^{(n)}}) =0$ at $\theta_H=0$.
The values $I_{W^{(n)}}$ are plotted in Fig.~\ref{HWWyukawa} for  
$n_F=3$ and  $\theta_H= 0.360$ ($z_L=10^8$).
One sees that the sign of $I_{W^{(n)}}$ alternates as $n$ increases, and its magnitude 
is almost constant; 
$I_{W^{(n)}} \sim (-1)^n \{ 0.14 + 0.0025 \ln n + 0.0011 (\ln n)^2 \}$ 
in the range $50 < n < 200$.
Note that $|g_{HW^{(n)} W^{(n)}}|$ itself increases with $m_{W^{(n)}}$, 
in sharp contrast to the behavior in the UED models.\cite{ued}

Similar behavior is observed for the Yukawa couplings of the top tower. One finds that
\beqn
&&\hskip -1.cm
I_{t^{(n)}} = 
\frac{y_{t^{(n)}} }{ y_t^\SM \cos \theta_H }
= - \frac{g_w}{2 y_t^\SM} \sqrt{ kL (z_L^2-1) } ~  \frac{\sin\theta_H}{N_{t^{(n)}}} \, 
\frac{C_L (1; \lambda_{t^{(n)}}, c_t)}{S_L(1; \lambda_{t^{(n)}}, c_t)} ~, \cr
\noalign{\kern 10pt}
&&\hskip -1.cm
N_{t^{(n)}} = \int_1^{z_L} dz  
\Big\{ (1+ \cos^2 \theta_H + 2 r_t) C_L(z; \lambda_{t^{(n)}}, c_t)^2
+ \sin^2 \theta_H \hat S_L (z; \lambda_{t^{(n)}}, c_t)^2 \Big\} ~, \cr
\noalign{\kern 10pt}
&&\hskip -.5cm
C_L(z; \lambda, c) = \frac{\pi}{2} \lambda \sqrt{z z_L}  
F_{c+(1/2),c-(1/2)}(\lambda z, \lambda z_L) ~, \cr
\noalign{\kern 10pt}
&&\hskip -.5cm
\hat S_L(z; \lambda, c) = \frac{C_L(1; \lambda, c) }{S_L(1; \lambda, c) } S_L(z; \lambda, c) ~.
\label{Htt1}
\eeqn
The values $I_{t^{(n)}}$ are plotted in Fig.~\ref{HWWyukawa}.   
The value of $I_{t^{(n)}}$ alternates in sign as $n$ increases, and the  magnitude 
of $y_{t^{(n)}}$  are almost  constant for large $n$.

The behavior of  the $y_{F^{(n)}}$ of the extra fermion is slightly different. 
In contrast to quarks and leptons, the lowest mode $F^{(1)}$ in the KK tower of $\Psi_F$ 
is massive at $\theta_H=0$; its mass is approximately given by 
$m_{F^{(1)}}(\theta_H) \propto \cos \onehalf \theta_H$.
Its Yukawa coupling is, therefore, expected to be  
$y_{F^{(1)}} \propto \sin \onehalf \theta_H$, becoming small for
small $\theta_H$.  Indeed one finds 
\beqn
&&\hskip -1.cm
I_{F^{(n)}} = \frac{y_{F^{(n)}}}{y_t^\SM \sin \onehalf \theta_H} =
- \frac{g_w}{4 y_t^\SM} \sqrt{ kL (z_L^2-1) } ~  \frac{\cos \onehalf \theta_H}{N_{F^{(n)}}} \, 
\frac{C_R (1; \lambda_{F^{(n)}}, c_F)}{S_R(1; \lambda_{F^{(n)}}, c_F)} ~, \cr
\noalign{\kern 10pt}
&&\hskip -1.cm
N_{F^{(n)}} = \int_1^{z_L} dz  
\bigg\{ \sin^2 \frac{\theta_H}{2} \,  C_R(z; \lambda_{F^{(n)}}, c_F)^2
+ \cos^2 \frac{ \theta_H}{2} \,  \hat S_R (z; \lambda_{F^{(n)}}, c_F)^2 \bigg\} ~, \cr
\noalign{\kern 10pt}
&&\hskip -.5cm
\hat S_R(z; \lambda, c) = \frac{C_R(1; \lambda, c) }{S_R(1; \lambda, c) } S_R(z; \lambda, c) ~.
\label{HFF1}
\eeqn
$I_{F^{(n)}}$ is plotted in Fig.~\ref{HWWyukawa}.
Again the value of $y_{F^{(n)}}$ alternates in sign as $n$ increases.  
For large $n$, $y_{F^{(n)}}/ y_t^\SM  \sim (-1)^n \, 0.14$  for $(n_F, z_L) = (3,10^8)$.

With all these Higgs couplings at hand, one can evaluate the rate 
$\Gamma (H \go \gamma \gamma)$ in (\ref{Hdecay1}).  
As all KK masses increase as $m_n \sim \onehalf n \, m_\KK $ for large $n$ and 
$F_1, F_{1/2}$ approach constant, the infinite sums rapidly  converge as 
$\sum (-1)^n n^{-1}$ or $\sum (-1)^n n^{-1} (\ln n)^q$ ($q=1, 2, \cdots$).
We have found that the sums saturate with about 50 terms.
The behavior of the alternating sign in the Higgs couplings of KK states has been
previously noticed in Refs.~\cite{MaruOkada} and \cite{Falkowski2008}.
Let $\cF_{W\,{\rm only}}$ ($\cF_{{\rm top ~ only}}$)  be the contribution of 
$W = W^{(0)}$ ($t=t^{(0)}$) to $\cF_W$ ($\cF_{\rm top}$) in (\ref{Hdecay1}).
For $n_F=3$ and $z_L= 10^8 \, (10^5)$, which yields $\theta_H =0.360\,  (0.117)$, 
one finds that 
$ \cF_{W\,{\rm only}} = 7.873 \, (8.330)$, 
$\cF_{\rm top ~ only} = - 1.305 \, (-1.372)$, 
$\cF_W/ \cF_{W\,{\rm only}} = 0.998 \, (0.9996)$,  
$\cF_{\rm top}/ \cF_{\rm top ~only} = 0.990 \, (0.998)$ and  
$\cF_F/ \cF_{\rm top ~only} = - 0.033 \, (-0.0034)$.

Let us suppose $Q_X^{(F)}=0$. 
In this case we obtain $\cF_W + \frac{4}{3} \cF_{\rm top} + \frac{3}{2} \cF_F = 6.199 \, (6.508)$.
Its ratio to $\cF_{W\,{\rm only}} + \frac{4}{3}\cF_{\rm top ~only}$ is 1.011 (1.001).
In other words, the contributions from KK states and $\Psi_F$ amount to only 1 \% (0.1 \%).
The dominant effect for the decay rates comes from the $\cos \theta_H$ suppression factor
in the amplitudes.  Compared to the value in the SM, $\Gamma [H \go \gamma \gamma ]$
is suppressed by 10 \% (1 \%) for $\theta_H = 0.360 \, (0.117)$.
The decay rate to two gluons is \cite{Georgi1978, Dawson1991, Spira1995}
\beeq
\Gamma ( H \go gg ) = \frac{\alpha_s^2 g_w^2 m_H^3}{128 \pi^3 m_W^2}
\, \big|  \cF_{\rm top}  \big|^2 ~,  
\label{Hdecay2}
\eneq
if $\Psi_F$ is a color singlet.  If $\Psi_F$ is a color triplet, $\cF_{\rm top}$ is replaced by
$\cF_{\rm top} + \frac{2}{3} n_F \cF_F$ in the above formula.
The correction due to KK excited states is small.

In the gauge-Higgs unification all decay rates for 
$H \go WW, ZZ, c \bar c, b \bar b, \tau \bar \tau$
are suppressed by a common factor $\cos^2 \theta_H$ at the tree level.
We have found that loop corrections due to KK excited states are very small.
Consequently the correction to the branching fraction of $H \go \gamma \gamma$ 
turns out very small, about 2\% (0.2\%) in the gauge-Higgs unification 
for $\theta_H = 0.360 \, (0.117)$.
The observed event rate for $H \go \gamma \gamma$, for instance, is determined by
the product of the Higgs production rate and the branching fraction, 
$\sigma_H^{\rm prod} \cdot B(H \go \gamma \gamma)$.
The production rate is suppressed, compared to the SM, by $\cos^2 \theta_H$,
but the branching fractions remain nearly the same as in the SM.
The gauge-Higgs unification predicts that the signal strength relative to the SM 
is   $\sim \cos^2 \theta_H$.  For $\theta_H = 0.1 \, (0.3)$, it is about 0.99 (0.91).
This is in sharp contrast to other models.  
In the UED models the contributions of KK states to $\cF_{\rm top}$ add up in the same sign
and may become sizable.\cite{ued}
In the gauge-Higgs unification  the contributions alternate in sign in the amplitudes,
resulting in the destructive interference and giving very small correction.

The rate $\Gamma(H\to \gamma\gamma)$ in Eq.~(4) can be enhanced through the factor $2 (Q_X^{(F)})^2+\onehalf$ for sufficiently  large $Q_X^{(F)}$.
For example, for $Q_X^{(F)}=4$ and $n_F=3$, we obtain the enhancement by a factor 2.22 (1.13) compared with the SM. 

The fact $m_H \sim 126\,$GeV leads to important consequences  in the gauge-Higgs unification.
We have found the universal relations  among  $m_\KK, \lambda_3, \lambda_4$ 
and $\theta_H$, which are independent of how many extra fermions are introduced.
The low energy data, the $S$ parameter constraint, and the tree-level unitarity
constraint indicate small $\theta_H < 0.3$.  The KK mass scale $m_\KK$ is 
predicted to be $3 \sim 7 \,$TeV for $\theta_H = 0.1 \sim 0.3$.
The existence of new charged heavy particles  
can affect the production and decay rates of the Higgs boson through loop diagrams.  
There are many proposals of models which employ such a mechanism to predict
the enhancement of the $H \go \gamma \gamma$ mode over other decay 
channels.\cite{ued}-\cite{ext-higgs}
In the gauge-Higgs unification there are new charged heavy particles, namely KK 
excited states of $W$ and top quark.  However, we have shown that their couplings
to the Higgs boson alternate in sign in each KK tower so that the correction
to the decay and production rates becomes very small. The gauge-Higgs unification
gives phenomenology at low energies very close to that of the SM so long as
$Q_X^{(F)}$ is moderately small.

Nevertheless  new rich structure is predicted to emerge.
We have seen above that the  cubic and quartic self-couplings of the Higgs boson significantly 
deviate from the SM.  The most clear signal for the gauge-Higgs unification would be
the production of the first KK states of the $Z$ boson and photon at LHC.
Their masses are predicted, for $\theta_H = 0.117 ~ (0.360)$, to be
$m_{Z^{(1)}} = 5.910 ~ (2.414)\,$TeV and $m_{\gamma^{(1)}} = 5.913~ (2.421)\,$TeV.
The current data \cite{ATLAS-Z1}-\cite{CMS-Z2}  indicate $m_{Z^{(1)}} > 2.5\,$TeV.
We have checked that there is a universal relation between $\theta_H$ and $m_{Z^{(1)}}$,
independent of $n_F$.  The data therefore imply that $\theta_H < 0.35$.
Another robust signal would be the production of a pair of the first KK state 
of the extra fermion,  $F^{(1)} \bar F^{(1)}$, which become stable. 
So far no new exotic stable charged fermion has been observed at LHC.\cite{CMS-F}
Its current limit puts a constraint $m_{F^{(1)}} > 0.5\,$TeV.
The value of $m_{F^{(1)}}$ depends on both $\theta_H$ and $n_F$ so that
no universal relation between $m_{F^{(1)}}$ and $\theta_H$ is found.
$m_{F^{(1)}}$ becomes smaller as $n_F$ increases with $\theta_H$ fixed.
$m_{F^{(1)}} > 0.5\,$TeV  implies   $\theta_H < 0.45$  for $n_F=3$.
We will come back to these issues with more details separately.

\vskip .5cm

\noindent
{\bf Acknowledgement:} 

This work was supported in part 
by  scientific grants from the Ministry of Education and Science, 
Grants No.\ 20244028, No.\ 23104009 and  No.\ 21244036.

\vskip .5cm

\ignore{
\renewenvironment{thebibliography}[1]
         {\begin{list}{[$\,$\arabic{enumi}$\,$]}  
         {\usecounter{enumi}\setlength{\parsep}{0pt}
          \setlength{\itemsep}{0pt}  \renewcommand{\baselinestretch}{1.2}
          \settowidth
         {\labelwidth}{#1 ~ ~}\sloppy}}{\end{list}}
}

\def\jnl#1#2#3#4{{#1}{\bf #2} (#4) #3}

\def\Zphys{{\em Z.\ Phys.} }
\def\jssc{{\em J.\ Solid State Chem.\ }}
\def\jpsJ{{\em J.\ Phys.\ Soc.\ Japan }}
\def\ptps{{\em Prog.\ Theoret.\ Phys.\ Suppl.\ }}
\def\PTP{{\em Prog.\ Theoret.\ Phys.\  }}
\def\JMP{{\em J. Math.\ Phys.} }
\def\NPB{{\em Nucl.\ Phys.} B}
\def\NP{{\em Nucl.\ Phys.} }
\def\PLB{{\it Phys.\ Lett.} B}
\def\PL{{\em Phys.\ Lett.} }
\def\PRL{\em Phys.\ Rev.\ Lett. }
\def\PRB{{\em Phys.\ Rev.} B}
\def\PRD{{\em Phys.\ Rev.} D}
\def\PRe{{\em Phys.\ Rep.} }
\def\AP{{\em Ann.\ Phys.\ (N.Y.)} }
\def\RMP{{\em Rev.\ Mod.\ Phys.} }
\def\ZPC{{\em Z.\ Phys.} C}
\def\SCI{\em Science}
\def\CMP{\em Comm.\ Math.\ Phys. }
\def\MPLA{{\em Mod.\ Phys.\ Lett.} A}
\def\IJMPA{{\em Int.\ J.\ Mod.\ Phys.} A}
\def\IJMPB{{\em Int.\ J.\ Mod.\ Phys.} B}
\def\EPJC{{\em Eur.\ Phys.\ J.} C}
\def\PR{{\em Phys.\ Rev.} }
\def\JHEP{{\em JHEP} }
\def\JCAP{{\em JCAP} }
\def\cmp{{\em Com.\ Math.\ Phys.}}
\def\JPA{{\em J.\  Phys.} A}
\def\JPG{{\em J.\  Phys.} G}
\def\NJP{{\em New.\ J.\  Phys.} }
\def\CQG{\em Class.\ Quant.\ Grav. }
\def\ATMP{{\em Adv.\ Theoret.\ Math.\ Phys.} }
\def\ibid{{\em ibid.} }

\renewenvironment{thebibliography}[1]
         {\begin{list}{[$\,$\arabic{enumi}$\,$]}  
         {\usecounter{enumi}\setlength{\parsep}{0pt}
          \setlength{\itemsep}{0pt}  \renewcommand{\baselinestretch}{1.0}
          \settowidth
         {\labelwidth}{#1 ~ ~}\sloppy}}{\end{list}}

\def\reftitle#1{{\it ``#1'' }}    


\end{document}